\documentclass[twocolumn,showpacs,preprintnumbers,amsmath,amssymb]{revtex4}

\usepackage{graphicx}
\usepackage{dcolumn}
\usepackage{bm}
\usepackage{graphics}
\usepackage{amsmath}
\usepackage{amssymb}
\usepackage{amscd}
\usepackage{afterpage}
\usepackage{float,times}
\usepackage{subfigure}
\usepackage{rotating}
\usepackage{multirow}
\usepackage{fancyheadings}
\usepackage{epsfig}
\usepackage{theorem}
\usepackage{moreverb}
\usepackage{euscript}
\usepackage{psfrag}

\begin{document}

\title{The Type-II Singular See-Saw Mechanism}

\author{K. L. McDonald}\email{k.mcdonald@physics.unimelb.edu.au}

\author{B. H. J. McKellar}%
 \email{b.mckellar@physics.unimelb.edu.au}
\affiliation{%
School of Physics, Research Centre for High Energy Physics, The
University of Melbourne, Victoria, 3010, Australia\\
}%


\begin{abstract}
The singular see-saw mechanism is a variation of the see-saw mechanism
whereby the right-chiral neutrino Majorana mass matrix is
singular. Previous works employing the singular see-saw mechanism have assumed
a vanishing left-chiral Majorana mass matrix. We
study the neutrino spectrum obtained under a singular see-saw
mechanism when the left-chiral neutrinos possess a non-zero Majorana
mass matrix. We refer to this as the type-II singular see-saw
mechanism. The resulting neutrino spectrum is found to be sensitive to
the hierarchy of the Dirac and Majorana mass scales used and
we explore the phenomenological consequences of the candidate
hierarchies. The compatibility of the resulting spectra with the body
of neutrino oscillation data is discussed. It is found that neutrino
mass matrices with this structure result in $3+1$ or $2+2$
neutrino spectra, making it
unlikely that this mass matrix structure is realized in nature. If the
left-chiral Majorana mass matrix is also singular we show that a
type-II singular see-saw mechanism can realize a spectrum of one
active-sterile pseudo-Dirac neutrino in conjunction with two active
Majorana neutrinos effectively decoupled from the sterile sector. This
realizes a scheme discussed in the literature in relation to
astrophysical neutrino fluxes.
\end{abstract}

\maketitle

\section{\label{sec:intro}INTRODUCTION\protect\\}
Over recent years our understanding of neutrino physics, in particular
their masses and mixings, has dramatically increased (for a review,
see for
example~\cite{king_review,giunti_laveder,smirnov_neutrino_review}).
The solar~\cite{solar} and
atmospheric~\cite{atmospheric} data, together with the terrestrial experiments KamLand~\cite{kamland} and K2K~\cite{k2k}, can be adequately accommodated by active neutrino oscillations. The bounds on $|U_{e3}|$ obtained by the reactor experiments CHOOZ~\cite{chooz} and Palo Verde~\cite{palo_verde} mean that the solar and atmospheric neutrino oscillations are practically decoupled~\cite{bilenky_giunti_solar_atm_decoupled}, whilst
the reconciliation of the LSND~\cite{lsnd} result with the other data remains
puzzling. Currently favoured fits imply bounds on the sterile component of active flavours
produced in the sun and the atmosphere that are increasingly
stringent (though questions regarding atmospheric
$\nu_{\mu}\rightarrow\nu_{\tau}$ verses $\nu_{\mu}\rightarrow\nu_{s}$
distinction~\cite{foot_superk_multiring} and the use of the $2+2$ sum
rule~\cite{pas_song_weiler} have been asked). Nevertheless there exist
regions of
the ($\sin^{2}2\theta , \delta m^{2}$) plane not yet experimentally
probed and the possibility remains that sterile neutrinos play a role in these regions,
independent of any role played in current experiments. 

The singular see-saw mechanism is a variation of the see-saw mechanism
whereby the right-chiral neutrino Majorana mass matrix is singular. In
previous works, where the singular see-saw mechanism was introduced to
accommodate the LSND result, the left-chiral neutrino Majorana mass
matrix was assumed to vanish~\cite{chun_kim_lee}. In this paper
we extend the singular see-saw mechanism by
including a left-chiral Majorana mass matrix. This is analogous to the type-II see-saw mechanism whereby the see-saw mechanism is extended to include a left-chiral Majorana mass matrix. The resulting neutrino spectrum is found to be sensitive to the hierarchy of the Dirac and Majorana mass scales used and
we explore the phenomenological consequences of the candidate hierarchies. We find that neutrino mass matrices with this structure can produce both $3+1$ and $2+2$ spectra, but as these spectra have difficulties accommodating all the oscillation data it is unlikely that this mass matrix structure is realized in nature. 

We also show that if the left-chiral Majorana mass matrix is singular
the type-II singular see-saw mechanism can give rise to an
active-sterile pseudo-Dirac pair of Majorana neutrinos. This realizes
a scheme recently discussed in the literature where one or more of the
mass eigenstates that participate in the solar and atmospheric
neutrino oscillations forms a pseudo-Dirac neutrino with a near
degenerate sterile
neutrino~\cite{bell_pseudo_dirac_neutrinos,keranen_maalampi_one_pseudo_dirac}.

The structure of this paper is as follows. In Section
\ref{sec:the_seesaw_mechanism} we review the see-saw mechanism in both
its standard and type-II (non-canonical) form. Section \ref{singular_seesaw} contains a discussion of the singular see-saw mechanism. The equivalent of a type-II form for the singular see-saw mechanism is introduced in Section \ref{nonzero_active_maj_mass} and the phenomenological consequences of the relevant scale hierarchies are compared with the oscillation data. In Section \ref{detecting_pseudo_dirac_pair} we discuss some of the  phenomenology resulting from one active-sterile pseudo-Dirac pair with very small mass splitting. 
\section{The See-Saw Mechanism\label{sec:the_seesaw_mechanism}}
It is known~\cite{see_saw} that the relative lightness of the neutrinos
can be explained by employing the so called see-saw mechanism. In
its standard form, the neutrino mass matrix in the Majorana basis is
given by:  
\[ \mathcal{M} = \left( \begin{array}{cc}
           0         & M_{D}   \\
           M^{T}_{D} & M_{R}   
\end{array} \right), \]
where the Dirac mass matrix \mbox{$ M_{D}$} and the Majorana
mass matrix $M_{R}$ are \mbox{$n \times n$} matrices for $n$
generations. If the elements of $M_{R}$ are of order $M$ and the
elements of $M_{D}$ are of order $m$, with $M\gg m$, then
diagonalization produces $n$ light Majorana
neutrinos and $n$ heavy
Majorana neutrinos with masses of order
\mbox{$m_{\nu} \sim m^{2}/M$} and \mbox{$m_{N} \sim M$}
respectively. To first order the heavy eigenvalues are found by
diagonalizing $M_{R}$ whilst diagonalizing
\mbox{$\mathcal{M}_{Light}=-M_{D}^{T}M_{R}^{-1}M_{D}$} gives the light
eigenvalues. Typically one identifies the light Majorana neutrinos
with the active neutrinos observed in nature, whilst the scale $M$ is set
by new physics. When the heavy Majorana neutrinos are integrated out
the light neutrinos are described by an effective dimension-five
operator~\cite{dim_five}: 
\begin{eqnarray}
& &
\mathcal{L}_{eff}=\frac{1}{\Lambda}\phi^{o}\phi^{o}\overline{(\nu_{i})^{c}}\nu_{j}
\label{eff_op}
\end{eqnarray}
where $\phi^{o}$ is the standard model scalar and \mbox{$\Lambda^{-1}=g_{i}g_{j}/2M$}. $g_{i}$ is the Yukawa
coupling constant for the Dirac mass. The number of light
neutrinos that the see-saw mechanism can naturally accommodate is
determined by the rank of the matrix
$M_{R}$~\cite{ecker_grimus_gronau} and in particular if $M_{R}$ is
rank $n$ one may obtain $n$ naturally light neutrinos. Hence at most one
can naturally obtain three
generations of light neutrinos if three heavy sterile neutrinos
exist~\cite{four_light_seesaw}.

If a non-zero active Majorana mass matrix $M_{L}$ is present, the
see-saw formula is modified. The effective mass matrix in the
light-sector takes the form:
\begin{eqnarray}
M_{Light}\simeq M_{L}-M_{D}^{T}M_{R}^{-1}M_{D}.\label{type_two_seesaw}
\end{eqnarray}
This form has been referred to as the type-II see-saw
formula. In general the scale of the entries in $M_{L}$ are completely
independent from those in $M_{R}$. When $M_{L}\neq 0$ the question of which term is larger
in eq. (\ref{type_two_seesaw}) arises. If the see-saw term
$-M_{D}^{T}M_{R}^{-1}M_{D}$ dominates, the original motivation
for the see-saw mechanism remains, with the scale of the light
neutrinos set by the suppressing large scale of
$M_{R}$. Though the presence of $M_{L}$ in the type-II see-saw
means the question `Why are the
neutrinos so light?' is effectively rephrased as `Why is $M_{L}$ so
light?'. Alternatively, if $M_{L}$ dominates the see-saw mechanism no
longer plays any part in setting the light sector mass scale (for a
discussion of the possible connection between large
$\nu_{\mu}-\nu_{\tau}$ mixing and a dominant $M_{L}$ in a
type-II see-saw scenario within SO(10)
see~\cite{bajc_senjanovic_vissani}). The type-II see-saw with dominant
$M_{L}$ allows the
neutrinos to possess large Dirac mass terms which do not significantly
affect the light eigenvector structure. The existence of the
relatively light active
neutrinos is attributed to the light scale of $M_{L}$.
\section{The Singular See-Saw Mechanism\label{singular_seesaw}}
It should be noted that the approximations used to develop the see-saw
mechanism  rely on the existence of
$M_{R}^{-1}$. There is in fact a history of study into the neutrino spectrum when $M_{R}$ is
singular. Singular Majorana mass matrices were of interest in SO(10)
when it was realized that combining~\cite{gron_john_schecht} the
Stech~\cite{stech} and Fritsch~\cite{fritzsch} ansatz for the quark
mass matrices in an SO(10) framework~\cite{bottino_kim} could lead to
singular behaviour in the Majorana mass sector~\cite{JRS}. 

The apparent observation of a small admixture of a $\sim$17~keV
neutrino state in $\nu_{e}$~\cite{simpsons_neutrino} motivated studies
that could produce such a
state. The size of the claimed mass meant the properties of
the neutrino were constrained by both neutrino-less double $\beta$-decay
experiments and cosmological arguments and a (pseudo-)Dirac particle
seemed necessary to explain the data. The singular see-saw
mechanism~\cite{sssm_glashow,allen_john_etc}
received interest in this context as it could produce a pseudo-Dirac
particle with the required properties. Specific models that realized a
singular $M_{R}$ amended the standard model to include an extra Abelian
symmetry~\cite{babu_mohapatra_glashow_model}, a non-Abelian
symmetry~\cite{foot_king} and horizontal
symmetries~\cite{partially_broken_ssm,davidson}, whilst a supersymmetric model
was also developed~\cite{du_liu}.

The singular see-saw mechanism also received interest in light of
LSND~\cite{lsnd} and the three distinct $\Delta m^{2}$
scales required to simultaneously explain the solar data~\cite{solar},
the atmospheric data~\cite{atmospheric} and the LSND results in terms
of neutrino oscillations. A general analysis was carried out
in~\cite{chun_kim_lee} and the possibility of a hierarchy in the Dirac mass
matrix was considered in~\cite{dirac_hierarchy}. A model using Abelian
symmetries that simultaneously produced singularities in the Dirac
mass matrix and the sterile Majorana mass matrix was also
constructed~\cite{liu_song}. General discussions regarding the
coexistence of large active-active and large active-sterile mixing in
the presence of a singular Majorana mass matrix can
be found
in~\cite{mckellar_stephenson_goldman_garbutt}.

The singular see-saw mechanism is a variation of the original see-saw
mechanism in which an $n$-dimensional matrix $M_{R}$ has
rank $(n-1)$ (or less). Depending on the model, it may be possible to obtain four
relatively light Majorana neutrinos, for three generations, via the
singular see-saw mechanism for a
rank 2 matrix $M_{R}$. Only two of the light
neutrinos have a naturally light mass, whilst the lightness of the
other two requires a light Dirac mass matrix. To see this
we consider the example of three generations of active and sterile
neutrinos with a matrix $M_{R}$ of rank 2~\cite{allen_john_etc,chun_kim_lee}. We first proceed by
diagonalizing the matrix $M_{R}$ and placing the zero eigenvalue in
the one-one entry so that the full mass matrix now reads:
\[ \left( \begin{array}{cc}
           0         & M_{D}'   \\
           M'^{T}_{D} & M_{R}'   
\end{array} \right) \]
where $M_{R}'=RM_{R}R^{T}=$diag($0,M_{1},M_{2}$), $M_{D}'=M_{D}R^{T}$
and R is the rotation matrix used to diagonalize $M_{R}$. Now define
new matrices:
\[ \left( \begin{array}{cc}
           0         & M_{D}'   \\
           M'^{T}_{D} & M_{R}'   
\end{array} \right) 
=\left( \begin{array}{cc}
           M_{\omega} & M_{\gamma}   \\
           M_{\gamma}^{T} & M_{d}   
\end{array} \right) \equiv \mathcal{M}'\]
where $M_{d}=$diag($M_{1},M_{2}$) is a \mbox{$2\times2$} diagonal
matrix, $M_{\gamma}$ is a \mbox{$4\times2$} matrix and $M_{\omega}$ is
a \mbox{$4\times4$} matrix with the zero eigenvalue of the matrix
$M_{R}$ in the entry $M_{\omega 44}$. The general form of $M_{\omega}$
is
\begin{eqnarray} M_{\omega}=\left( \begin{array}{cccc}
           0     & 0 & 0 & a_{1}  \\
           0     & 0 & 0 & a_{2}  \\
           0     & 0 & 0 & a_{3}  \\
           a_{1} & a_{2} & a_{3} & 0   
\end{array} \right).\label{m_omega_defined} 
\end{eqnarray}
The eigenvalue equation may now be solved. It is given by:
\begin{eqnarray}
& & \mathcal{M}' \Omega = \lambda \Omega
\label{ev_eq_ssw}
\end{eqnarray}
where \mbox{$\Omega^{T}=( \psi^{T}_{Light}, \psi^{T}_{Heavy})$} and
$\psi_{Light}$ ($\psi_{Heavy}$) is a four (two) dimensional column
vector. Equation (\ref{ev_eq_ssw}) is equivalent to the two coupled
equations
\begin{eqnarray}
\label{ev_one}
& & M_{\omega}\psi_{Light} + M_{\gamma}\psi_{Heavy} = \lambda
\psi_{Light} \\
& & M_{\gamma}^{T}\psi_{Light} + M_{d}\psi_{Heavy}  = \lambda
\psi_{Heavy}.
\label{ev_two}
\end{eqnarray}
Solving (\ref{ev_two}) for $\psi_{Heavy}$ and substituting into
(\ref{ev_one}) gives
\begin{eqnarray}
& & (M_{\omega} -M_{\gamma}(M_{d}-\lambda)^{-1}M_{\gamma}^{T})\psi_{Light} =
\lambda \psi_{Light}
\label{bootstrap_ev}
\end{eqnarray}
which has the form \mbox{$f(\lambda ) \psi_{Light} = \lambda
\psi_{Light}$}. This equation may be solved to obtain the four lighter
eigenvalues. As we expect $\lambda\ll M$ for the light eigenvalues we
can use \mbox{$(M_{d}-\lambda)^{-1} \approx M_{d}^{-1}$}, giving:
\begin{eqnarray}
& & (M_{\omega} -M_{\gamma}M_{d}^{-1}M_{\gamma}^{T})\psi_{Light} =
\lambda \psi_{Light}
\label{light_ev}
\end{eqnarray}
as the light sector eigenvalue equation to first order. The zeroth
order eigenvalues are obtained by solving
\begin{eqnarray*}
& & M_{\omega}\psi_{Light} =
\lambda \psi_{Light}
\end{eqnarray*}
and we see that the scale of the mass eigenvalues is set by the
eigenvalues of $M_{\omega}$, which are
\mbox{$\lambda^{(0)}=\left\{0,0,m_{\omega},-m_{\omega}\right\}$} with
\mbox{$m_{\omega}=(a_{1}^{2}+a_{2}^{2}+a_{3}^{2})^{1/2}$}. We expect
the \mbox{$a_{i}$'s} to be O($m$), where $m$ is the scale of the neutrino Dirac mass matrix, so
that two of the eigenvalues are ultra light whilst the other two have
a mass set by the Dirac scale. It is worth noting that $M_{\omega}$
possesses a lepton number symmetry
$\tilde{L}=L_{e}+L_{\mu}+L_{\tau}+L_{s}$, as the only non-zero entries
are Dirac mass terms coupling the active neutrinos to the massless
eigenvalue of $M_{R}$, $\nu_{s}$. Consequently the lowest order
eigenstates must be Dirac particles or massless Majorana (Weyl)
states~\cite{wolfenstein_massive_dirac_neutrinos}. The term
\mbox{$M_{\gamma}M_{d}^{-1}M_{\gamma}^{T}$} provides corrections to the
eigenvalues of order $m^{2}/M$. It will break $\tilde{L}$ and split
the Dirac particle to a pseudo-Dirac pair. The degeneracy of the
massless particle will in general be lifted and a pair of massive
Majorana neutrinos results. The two intermediate scale eigenvectors are:
\begin{eqnarray}
\nu_{3,4}=\frac{1}{\sqrt{2}}\left(
  \frac{a_{1}\nu_{e}+a_{2}\nu_{\mu}+a_{3}\nu_{\tau}}{\sqrt{a_{1}^{2}+a_{2}^{2}+a_{3}^{2}}} \pm
  \nu_{s}\right)\label{intermediate_mass_estates}
\end{eqnarray}
and the light sterile field is seen to reside predominantly in the
pseudo-Dirac pair. The emergence of the pseudo-Dirac pair
justified the interest in the singular see-saw scenario in view of
Simpson's 17~keV neutrino results. In the more modern context, the
pseudo-Dirac pair produced near maximal mixing between an active
linear combination and the light sterile state. If the active component of
the pseudo-Dirac pair is taken as mostly $\nu_{e}$ ($\nu_{\mu}$) then
near maximal oscillations between $\nu_{e}$ ($\nu_{\mu}$) and $\nu{_s}$
occur.

Eq. (\ref{light_ev}) looks similar to the type-II see-saw formula
(\ref{type_two_seesaw}) but an
important difference exists. In the type-II formula the scales of
$M_{L}$ and the see-saw term $M_{D}^{T}M_{R}^{-1}M_{D}$ are
independent and the question of relative size arises. In the singular
see-saw mechanism the first term $M_{\omega}$ and the
see-saw type term $M_{\gamma}M_{d}^{-1}M_{\gamma}^{T}$ are related and the
first term always dominates. Thus the gross structure of the
eigenstates obtained via the singular see-saw mechanism is set
by $M_{\omega}$, giving a pseudo-Dirac pair which contain a large sterile
component and a pair of lighter Majorana neutrinos with a mass
suppressed by the large sterile Majorana scale.


\section{Non-Zero Active Majorana Mass\label{nonzero_active_maj_mass}}
We now consider the singular see-saw mechanism with a non-zero
Majorana mass matrix for the active
neutrinos. The equivalent of a type-II form for the singular
see-saw mechanism follows immediately. 
The mass matrix is
\[ \mathcal{M}= \left( \begin{array}{cc}
           M_{L}      & M_{D}  \\
           M^{T}_{D}  & M_{R}    
\end{array} \right)=O^{T}\left( \begin{array}{cc}
           UM_{L}U^{T}        & UM_{D}R^{T}  \\
           (UM_{D}R^{T})^{T}  & RM_{R}R^{T}    
\end{array} \right)O, \]
where $O=\mathrm{diag}(U,R)$ is a $6\times 6$ orthogonal matrix such
that $U$ ($R$) is the $3\times 3$ sub-matrix that digonalizes
$M_{L(R)}$. As we are taking $M_{R}$ to be singular we choose $R$ such
that $RM_{R}R^{T}=\mathrm{diag}(0,M_{1},M_{2})$ and we have
$UM_{L}U^{T}=\mathrm{diag}(m_{1},m_{2},m_{3})$. Defining:
\[ \tilde{M}_{L}\equiv \left( \begin{array}{cc}
           UM_{L}U^{T}        & 0_{3\times 1}  \\
           0_{3\times 1}^{T}  & 0    
\end{array} \right)=\mathrm{diag}(m_{1},m_{2},m_{3},0), \]
with $0_{3\times 1}^{T}=(0,0,0)$, we repartition the mass matrix as:
\[ \mathcal{M}=O^{T}\left( \begin{array}{cc}
           \tilde{M}_{L}+M_{\omega} & M_{\gamma}  \\
           M_{\gamma}^{T}           & \tilde{M}_{R}    
\end{array} \right)O.\label{partitioned_with_active_mass} \]
We have introduced $\tilde{M}_{R}\equiv\mathrm{diag}(M_{1},M_{2})$ and
the $4\times 2$ matrix $M_{\gamma}$. The matrix $M_{\omega}$ is a
$4\times 4$ matrix with
the same form as eq. (\ref{m_omega_defined}) and contains the zero
eigenvalue of $M_{R}$ as its (4,4) element. We denote the scale of the
non-zero elements in the Majorana and Dirac mass matrices as
$O(M_{L})\sim\tilde{m}$, $O(M_{R})\sim M$ and $O(M_{D})\sim m$. Taking
$\tilde{m}\ll M$ and $m\ll M$ allows us to block diagonalize the mass
matrix in (\ref{partitioned_with_active_mass}) up to $O(m^{2}/M)$ as:
\begin{eqnarray}
& &\left( \begin{array}{cc}
           \tilde{M}_{L}+M_{\omega} & M_{\gamma}  \\
           M_{\gamma}^{T}           & \tilde{M}_{R}    
\end{array} \right)\nonumber\\
&=&\left( \begin{array}{cc}
           1      & S  \\
           -S^{T} & 1    
\end{array} \right)\left( \begin{array}{cc}
           M_{\nu}^{eff} & 0  \\
           0             & \tilde{M}_{R}    
\end{array} \right)\left( \begin{array}{cc}
           1     & -S  \\
           S^{T} & 1    
\end{array} \right)\label{block_diag_active_case} 
\end{eqnarray}
with $S=M_{\gamma}\tilde{M}_{R}^{-1}$ and:
\begin{eqnarray}
M_{\nu}^{eff}=\tilde{M}_{L}+M_{\omega}-M_{\gamma}\tilde{M}_{R}^{-1}M_{\gamma}^{T}.\label{non_canonical_singular_seesaw} 
\end{eqnarray}
We refer to (\ref{non_canonical_singular_seesaw}) as the type-II
singular see-saw formula. The distinction between the
matrices $\tilde{M}_{L}$ and $M_{\omega}$
has been maintained as the scale of their non-zero entries will in
general be independent. As with the type-II see-saw formula, the
question of which matrix sets the scale for the light neutrino sector
arises. Though now it is a comparison of the Dirac scale and the
active Majorana scale that is relevant. We consider the two cases.
\subsection{Dirac Scale Dominance}
If we take $m \gg \tilde{m}$ then $M_{\omega}$ will determine the structure of
the four lightest eigenstates. This case is essentially the same as
the singular see-saw case. To lowest order we obtain two massless
states and a Dirac particle with mass $m_{\omega}$ as defined in
Section \ref{singular_seesaw}. The active Majorana mass terms will now
contribute to the splitting of the massless states and the Dirac
particle. If the the active mass scale is larger than the see-saw
correction $M_{\gamma}\tilde{M}_{R}^{-1}M_{\gamma}^{T}$, the size of the
splittings will differ from the canonical singular see-saw
case. Otherwise the neutrino spectrum of the singular see-saw case is
reproduced.
\subsection{Non-Negligible Active Majorana Mass}
The eigenvector structure is quite different when the effects of the
active Majorana mass matrix must be included. The lack of knowledge of
the overall neutrino mass scales necessitates a discussion of the
different viable alternatives. If the eigenvalues of $M_{L}$ are all
of order $\tilde{m}$ and  $\tilde{m} \gg m$, the gross structure of
the eigenstates is set by
the active Majorana mass matrix. To lowest order the light mass
eigenstates will be three purely active Majorana neutrinos with masses
$m_{i}$, $i=1,2,3$ , corresponding to the mass eigenstates of $M_{L}$,
and one massless sterile state. The non-zero
elements in $M_{\omega}$ will mix these states. In general each of the
purely active states will develop a small sterile component, whilst
the sterile state will acquire a corresponding active
component. Denoting the eigenvalues of $M_{L}$ as $\nu_{i}^{(0)}$,
$i=1,2,3$, and the massless eigenvector of $M_{R}$ as $\nu_{4}^{(0)}$,
we consider $M_{\omega}$ as a perturbation on 
$\tilde{M}_{L}$. The perturbed eigenvectors are
\begin{eqnarray}
\nu_{i}^{(1)}=\nu_{i}^{(0)}+\frac{a_{i}}{m_{i}}\nu_{4}^{(0)},\nonumber\\
\nu_{s}^{(1)}=\nu_{4}^{(0)}-\sum_{i=1}^{3}\frac{a_{i}}{m_{i}}\nu_{i}^{(0)},
\end{eqnarray}
where we have labelled $M_{\omega}$ as in
eq. (\ref{m_omega_defined}). The eigenvalues do not shift to first
order. The predominantly sterile state develops a non-zero mass at
second order, given by
\begin{eqnarray*}
m_{4}=\sum_{i=1}^{3}\frac{a_{i}^{2}}{m_{i}^{2}}.
\end{eqnarray*}
In this case the type-II singular see-saw gives rise to three predominantly active Majorana
neutrinos and one lighter predominantly
sterile Majorana neutrino. If one was to attempt to accommodate LSND
with such a mechanism the resulting spectrum would be classified as a
$3+1$ scenario. If $O(a_{i}/m_{i})\sim 0.1$ and $\tilde{m}\sim
1$~eV the resulting $3+1$ spectrum can accommodate the atmospheric and
solar data in
terms of oscillations amongst the active states $\nu^{(1)}_{i}$, with large
angle mixing built into the rotation $U$, which
diagonalizes $M_{L}$. Such a spectrum may not be able to explain the
solar, atmospheric and LSND data, with recent analysis suggesting a best overall goodness of fit for $3+1$
spectrums to be $5.6\times
10^{-3}$~\cite{maltoni_schwetz_tortola_valle}. Should the future
interpretation of the LSND result not require neutrino oscillations
the scale of $\tilde{m}$ can be much lower. The
sterile component in $\nu^{(1)}_{i}$ may be small
enough to comply with all other experimental constraints. This
component produces small amplitude oscillations into the sterile state
with the relevant mass-squared
differences for oscillations between the
\mbox{$\nu_{i}^{(1)}$'s} and $\nu_{4}^{(1)}$ set by the the the active mass scales, $\Delta
m^{2}_{i4}=m_{i}^{2}-m_{4}^{2}\approx m_{i}^{2}$. A current bound of $\sin^{2}\eta <0.52$ at
$3\sigma$ for
$\nu_{e}\rightarrow \cos\eta\nu_{\alpha} +\sin\eta\nu_{s}$, with
$\nu_{\alpha}$ containing only active
flavours, has been derived for sterile mixing with solar
neutrinos~\cite{bahcall_gonzalez-garcia_pena-garay}. An atmospheric
bound of $\sin^{2}\xi<0.19$ at the 90\% C.L. for
$\nu_{\mu}\rightarrow\cos\xi\nu_{\tau}+\sin\xi\nu_{s}$ has also been obtained~\cite{sterile_atm_bound}.

We have taken the eigenvalues of $M_{L}$ to be at least of order
$\tilde{m}$. This can be the case for the quasi-degenerate hierarchy ($m_{1}\simeq
m_{2}\simeq m_{3}$), the normal hierarchy ($m_{1}\ll m_{2}\ll m_{3}$)
and the inverted hierarchy ($m_{2}\gtrsim m_{1}\gg m_{3}$), though it
need not be an accurate assumption. The depletion
of solar and atmospheric neutrinos due to oscillations leads to lower
bounds for two of the mass eigenstates:
\begin{eqnarray*}
& &m_{i}\geq (\Delta m^{2}_{atm})^{1/2}\approx 5\times 10^{-2}\mathrm{eV},\\
& &m_{j}\geq (\Delta m^{2}_{\odot})^{1/2}\approx 7\times 10^{-3}\mathrm{eV},
\end{eqnarray*}
where the value attributed to $i,~j$ depends on the mass pattern. For
the normal and inverted hierarchies the lightest mass is unconstrained
and may have a vanishingly small value. Thus we can ask what happens to the type-II
singular see-saw spectrum if $M_{L}$ is also singular? This may be
expected in, for example, a left-right symmetric model. In this case we
have $0\simeq m_{1}<a_{i}\sim m\ll m_{2},m_{3}\sim \tilde{m}$. The
lowest order eigenvalues are obtained by diagonalizing the matrix
$\tilde{M}_{L}+M_{\omega}$, where now
$\tilde{M}_{L}=\mathrm{diag}(0,m_{2},m_{3},0)$. To order $m/\tilde{m}$
the eigenvalues are $\lambda^{0}=\{\pm a_{1},m_{2},m_{3}\}$ and the
sterile state is seen to form a Dirac particle with the zero
eigenvalue of $M_{L}$, corresponding to the non-standard lepton number
symmetry $L_{1}+L_{s}$ (where
$L_{1}$ is a lepton number given to $\nu_{1}$) present in
$\tilde{M}_{L}+M_{\omega}$ when $a_{2,3}\rightarrow 0$. Higher order
corrections to the mass matrix
will split the Dirac particle to form a pseudo-Dirac pair of Majorana
neutrinos. The interpretation of this spectrum depends on whether one
attempts to accommodate the LSND result or not.

The spectrum may be of interest if the LSND result is not explained in
terms of neutrino oscillations. The atmospheric and solar data would
be accommodated
by oscillations
between $\nu_{2}$, $\nu_{3}$ and the active component of
the light pseudo-Dirac pair, $\nu_{1}$. The mixing between members of the
pseudo-Dirac pair is near maximal, with the oscillation length
dependent on the size of their splitting. Provided the
ratio $m/\tilde{m}$ is small enough the oscillation
length of the pseudo-Dirac pair may be larger than solar system length
scales. At distances shorter than the oscillation length for the
pseudo-Dirac pair the presence of the sterile partner will not be
observable. Beyond the oscillation length a conversion of active
flavours into the sterile state occurs. Regardless of the size of the
role played by sterile neutrinos in solar and atmospheric neutrino
phenomena (if any at all), they may still play a role in regions of
the ($\sin^{2}2\theta , \delta m^{2}$) plane not yet probed. The smallest
mass-squared difference thus far probed is about
$10^{-11}\mathrm{eV}^{2}$ with solar neutrinos. If we take $m_{2}\sim
5\times 10^{-3}$, $m_{3}\sim 7\times 10^{-2}$ (corresponding to the
lowest bounds in a normal hierarchy) and $a_{i}\lesssim 10^{-5}$, the
splitting of the pseudo-Dirac pair is $\lesssim 10^{-11}$ and the
sterile partner of the lightest mass eigenstate would thus far have
escaped detection. With these values the mixing angle of the active
states $\nu_{2}$ and $\nu_{3}$ with the sterile state are $\sim
10^{-7}-10^{-8}$ and these states effectively decouple from the
sterile sector. The scale of the Dirac mass matrix required here is
small compared to the charged fermions, as is the case of the singular
see-saw mechanism. Though it is not our intention to build models we
note that mechanisms that induce small neutrino Dirac mass scales
exist. These include methods of excluding Dirac masses at tree
level~\cite{babu_mohapatra_glashow_model}, a Dirac
see-saw~\cite{chang_kong_dirac_seesaw} and mechanisms employing
large extra dimensions~\cite{small_dirac_by_extra_dimensions}.
\section{Detecting One Pseudo-Dirac Pair\label{detecting_pseudo_dirac_pair}}
Observation of
neutrino fluxes from astrophysical sources can reveal the existence of
a pseudo-Dirac structure connecting active and sterile states. Over the huge path length from
astrophysical neutrino sources the phases of the the relatively large
atmospheric and solar mass-squared differences effectively
decohere. The neutrino density matrix becomes mixed amongst the
active flavours whilst remaining coherent amongst pseudo-Dirac
partners. In the absence of pseudo-Dirac partners, the neutrino flux
from astrophysical sources is expected
to be flavour democratic~\cite{learned_pakvasa}. Pseudo-Dirac splittings
can lead to a flavour dependent oscillatory reduction in flux that is in
principle detectable. The effects on neutrino fluxes from
astrophysical sources with a sterile pseudo-Dirac partner for each active mass
eigenstate (a similar pattern to that occurring, for example, in the
mirror model~\cite{pseudo_dirac_mirror_model}) are discussed
in~\cite{bell_pseudo_dirac_neutrinos}.

A scenario with one sterile neutrino forming a near degenerate
mass pair with one active neutrino were considered
in~\cite{keranen_maalampi_one_pseudo_dirac}. The flux ratio of
ultra-high energy electron and muon neutrinos with the sterile
state present, $R^{sterile}_{e\mu}\equiv
\Phi^{sterile}_{e}/\Phi^{sterile}_{\mu}$, was compared with the value
predicted by the standard model for a range of mixing angles
between $\nu_{1}$ and $\nu_{s}$. Deviations between
$R^{sterile}_{e\mu}$ and $R^{SM}_{e\mu}$ were
found, permitting observation of a sterile pseudo-Dirac partner over
astrophysical length scales. The
deviations were most marked for near maximal mixing where there was no
overlap between the range of $R^{sterile}_{e\mu}$ and
$R^{SM}_{e\mu}$.

The recent improved determination of cosmological
parameters provides strong constraints on
the number of relativistic species present during Big Bang
Nucleosynthesis. Sterile neutrino populations created and maintained
by neutrino oscillations must have sufficiently small active-sterile
mixing and/or mass-squared differences to avoid disturbing the
standard nucleosynthesis of light elements. This leads to the bound
$|\delta m^{2}|\sin^{2}2\alpha <5\times 10^{-8}\mathrm{eV}^{2}$ for
$\nu_{e}\leftrightarrow \nu_{s}$ mixing~\cite{shi_fuller} whilst the
bounds for the other flavours are less severe. The value of $\delta
m^{2}<10^{-11}\mathrm{eV}^{2}$, taken to avoid disrupting solar
neutrino experiments, satisfies
this bound. Thus neutrino
oscillations involving one active-sterile pseudo-Dirac pair do not
create a significant sterile
population and the neutrino pattern predicted by the type-II singular
see-saw mechanism with a singular $M_L$ remains experimentally viable.
\section{Conclusion}
In this paper the singular
see-saw mechanism was extended to include a left-chiral Majorana mass
matrix. It was found that, depending on the hierarchy between the
Dirac and Majorana mass scales, the type-II singular see-saw predicted a $2+2$
or $3+1$ neutrino spectrum. As these spectra have difficulties in
accommodating the body of neutrino oscillation data it is unlikely
that this mass matrix structure is realized in nature. It was also shown
that a type-II singular see-saw produces a spectrum
containing one
active-sterile pseudo-Dirac pair and two active Majorana neutrinos if
the left-chiral
Majorana mass matrix $M_L$ is also singular.
\section*{Acknowledgements}
K.M. wishes to thank Catherine Low and Robert Foot for useful conversations. This work was supported in part by the Australian Research Council. 

\end{document}